\documentclass[10pt,showpacs,twocolumn]{revtex4-2}
\usepackage{color}
\usepackage{graphicx}
\usepackage{amsmath}
\usepackage{amssymb}
\usepackage{url}
\usepackage{braket}
\usepackage{physics}

\usepackage[
bookmarks=true,
colorlinks,
linkcolor=blue,
urlcolor=blue,
citecolor=blue,
plainpages=false,
pdfpagelabels,
final,
breaklinks=true
]{hyperref} 

\begin{document}

\preprint{APS/123-QED}
\title{Instantaneous optical selection rule for independent control of valley currents}
\author{Wanzhu He$^{1}$}
\author{Xiaosong Zhu$^{1}$}\email{zhuxiaosong@hust.edu.cn}
\author{Liang Li$^{1}$}\email{liangl@hust.edu.cn}
\author{Di Wu$^{1}$}
\author{Xiaotong Zhu$^{1}$}
\author{Pengfei Lan$^{1,2}$}\email{pengfeilan@hust.edu.cn}
\author{Peixiang Lu$^{1,2}$}\email{lupeixiang@hust.edu.cn }

\affiliation{%
$^1$Wuhan National Laboratory for Optoelectronics and School of Physics,
Huazhong University of Science and Technology, Wuhan 430074, China\\
$^2$Hubei Optical Fundamental Research Center, Wuhan 430074, China\\ 
}


\date{\today}

\begin{abstract}
We reveal an instantaneous optical valley selection rule that illuminates the coupling between the instantaneous optical chirality of the driving laser field and the chirality of valley systems. Building on this principle, 
we propose and demonstrate that a single chirality-separated optical field, in which oppositely signed instantaneous optical chiralities are separated within an optical cycle, enables independent manipulation of currents from $\rm K$ and $\rm K^{\prime}$ valleys.
Based on this scheme, we highlight two key example applications: (1) complete separation of currents from different valleys, yielding 100\%-purity valley-polarized currents, and (2) generation of pure valley current with zero net charge flow. Our work offers a robust and highly controllable all-optical strategy for ultrafast engineering valley currents in the optical cycle timescale, paving a new avenue for valleytronics and quantum information technologies. 
\end{abstract}

\maketitle

\textit{Introduction}---In two-dimensional (2D) hexagonal systems lacking spatial inversion symmetry, an additional degree of freedom, the valley degree of freedom (VDOF), emerges alongside charge and spin \cite{vitaleValleytronicsOpportunitiesChallenges2018a,makValleyHallEffect2014a,jiangMicrosecondDarkexcitonValley2018}. The band structure of these materials features two degenerate but inequivalent valleys $\rm K$ and $\rm K^{\prime}$, located at the energy extrema in the Brillouin zone (BZ). Due to the strong suppression of intervalley scattering, the valley index can be regarded as a pseudo-spin, and is largely conserved during electron transport. This unique property has fostered the development of valleytronics \cite{schaibleyValleytronics2DMaterials2016,degiovanniniMonitoringElectronPhotonDressing2016a,guptaObservation100Valleycoherent2023,yuValleyExcitonsTwodimensional2015,jimenez-galanSubcycleValleytronicsControl2021a,liAttosecondAllOpticalRetrieval2024a}, a field dedicated to utilizing VDOF to encode quantum information and control valley currents (currents generated by electrons from individual valleys).

 Two typical destinations of valley current manipulation are producing ``valley-polarized current'' (a total current predominantly contributed by one valley) and ``pure valley current'' (a total current with zero net charge flow but nonzero transport of pseudo-spin), which have attracted much attention over the past two decades.  Early approaches relied on rigid architectures, such as a fixed pair of electrodes to supply a strong external bias or organized heterostructures \cite{jinImagingPureSpinvalley2018b,settnesGrapheneNanobubblesValley2016,chenOneWayValleyRobustTransport2025,suiGatetunableTopologicalValley2015a,hsuNanoscaleStrainEngineering2020}. This precludes the possibility of ultrafast manipulation, as the timescale was limited by the inherent characteristics of these rigid structures. Besides, multi-pumped schemes combining a linearly polarized terahertz bias with resonant excitation pulses were also proposed, enabling manipulation on timescales from picoseconds to hundreds of femtoseconds \cite{sharmaTHzInducedGiant2023b,sharmaGiantControllableValley2023,gillUltrafastAllopticalGeneration2025a,langerLightwaveValleytronicsMonolayer2018a}. Recently, all-optical methods in the visible and mid-infrared regimes have been developed enabling ultrafast control of valley dynamics down to the optical cycle level \cite{mrudulLightinducedValleytronicsPristine2021a,tyulnevValleytronicsBulkMoS22024b,mitraLightwavecontrolledHaldaneModel2024a,oliaeimotlaghFemtosecondValleyPolarization2018,neufeldLightDrivenExtremelyNonlinear2021,motlaghAnomalousUltrafastAlloptical2021,ranaOpticalControlUltrafast2024,sharmaDirectCouplingLight2024a,liHighpurityValleypolarizedCurrents2025,leskoOpticalControlElectrons2025a,gillCreationControlValley2025}. 
 
The optical control of valleytronics usually exploits the optical valley selection rule (OVSR) \cite{yaoValleydependentOptoelectronicsInversion2008a,makControlValleyPolarization2012,geondzhianValleySelectivitySoft2022,herrmannNonlinearValleySelection2025b,hashmiUltrafastOpticalControl2025}, which arises due to the selective coupling between the inequivalent valleys and the cycle-averaged chirality of the optical field (e.g., the helicity of a circularly polarized field). 
The effect of OVSR leads to unequal residual excited populations in different valleys while interacting with chiral optical pulses. 
Meanwhile, nonvalishing valley currents can be achieved utilizing driving fields with asymmetric vector potentials.  However, in existing approaches, the electrons excited in both valleys experience the same asymmetric waveform of the driving field, and consequently, the charge carriers from both $\rm K$ and $\rm K^{\prime}$ valleys flow in the same direction \cite{oliaeimotlaghFemtosecondValleyPolarization2018,neufeldLightDrivenExtremelyNonlinear2021,motlaghAnomalousUltrafastAlloptical2021,ranaOpticalControlUltrafast2024,sharmaDirectCouplingLight2024a,liHighpurityValleypolarizedCurrents2025,leskoOpticalControlElectrons2025a,gillCreationControlValley2025}.
Namely, the control of the two valleys is synchronized, and the two VDOFs are not fully utilized, thereby limiting the flexibility of valleytronics applications. To overcome these limitations, a fundamental and robust solution requires independent control of valley currents simultaneously, which remains largely unexplored.

In this letter, we establish a theoretical framework that reveals an instantaneous OVSR, clarifying how instantaneous optical chirality (IOC) governs valley-resolved excitation dynamics. Guided by this mechanism, we propose that chirality-separated optical fields, with oppositely signed IOC separated within a cycle, enable independent control over both valleys simultaneously. We present a strategy for constructing such fields and demonstrate their capability to steer valley currents toward arbitrary directions independently. Moreover, we highlight two important applications: (1) complete separation of currents from different valleys, and (2) generation of pure valley current at the optical cycle timescale. 

\textit{Instantaneous optical valley selection rule} --- We consider a two-band 2D hexagonal system with a conduction band (CB) and a valence band (VB) exposed to a laser field, which has been extensively studied previously \cite{cuiValleyresolvedInterbandExcitation2022,oliaeimotlaghTopologicalResonanceSingleopticalcycle2019,parksBlochGaugeSymmetry2024,chaconCircularDichroismHigherorder2020b}, and investigate the OVSR via the CB population distribution $N_c({\mathbf{k}})$ in the BZ. The excitations can be determined by the temporal interference of the interband transition, which is described by the accumulated phase during a dynamical process $S\left[\mathbf{k}(t)\right]$ (See Sec.~$\rm{I}(A)$ in Supplemental Material (SM) for details \cite{hwzSM}):
\begin{equation}\label{phase}
\begin{aligned}
S\left\lbrack \mathbf{k}\!\left(t\right)\right\rbrack\!=&\!\int_{-\infty}^{t}dt^{\prime}E_{cv}\left\lbrack \mathbf{k}\!\left(t^{\prime}\right)\right\rbrack+\mathbf{F}\left(t^{\prime}\right)\cdot \mathbf{R}_{cv}\left\lbrack \mathbf{k}\!\left(t^{\prime}\right)\right\rbrack\\
&+{\rm Im}\left\{\frac{\dot{\mathbf{n}}(t^{\prime})\cdot \mathbf{d}_{cv}\left\lbrack \mathbf{k}\!\left(t^{\prime}\right)\right\rbrack}{\mathbf{n}(t^{\prime})\cdot \mathbf{d}_{cv}\left\lbrack \mathbf{k}\!\left(t^{\prime}\right)\right\rbrack}\right\},
\end{aligned}
\end{equation}
Here, $\mathbf{k}(t)=\mathbf{k}+\mathbf{A}(t)$ follows the acceleration theorem, with $\mathbf{A}(t)$ being the vector potential of the driving field. 
$E_{cv}=E_c-E_v$ is the energy difference between CB and VB. $\mathbf{R}_{cv}$ is known as the shift vector \cite{liPhaseInvarianceSemiconductor2019,qianRoleShiftVector2022}.
$\mathbf{F}\left(t\right)$ is the external field with the polarization vector $\mathbf{n}(t)=\mathbf{F}(t)/\|\mathbf{F}(t)\|$, and $\mathbf{d}_{cv}$ denotes the transition dipole moment between CB and VB.  

Notably, the third term of Eq.~(\ref{phase}) involves the time derivative $\dot{\mathbf{n}}(t)$, capturing the coupling between the system and the temporal variation of the polarization direction (thus the instantaneous rotation direction) of the optical field. This motivates us to introduce the IOC of the laser field $c(t)$ \cite{tangEnhancedEnantioselectivityExcitation2011,neufeldOpticalChiralityNonlinear2018,rozenControllingSubcycleOptical2019,note_c} and the chirality of the system  $\zeta(\mathbf{k})$  (See Sec.~$\rm{I}(B)$ in SM for details \cite{hwzSM})
\begin{align}
c(t)&=n_y(t)\partial_t{n}_x(t)-n_x(t)\partial_t{n}_y(t)\label{c}\\ 
\zeta(\mathbf{k})&={\rm Re}\left\{\frac{d_{-}(\mathbf{k})-d_{+}(\mathbf{k})}{d_{-}(\mathbf{k})+d_{+}(\mathbf{k})}\right\},
\end{align}
where $n_x$ and $n_y$ denote the $x$- and $y$-components of $\mathbf{n}(t)$ and $d_{\pm}$ represent the right- and left-handed components of the transition dipole moment $\mathbf{d}_{cv}$. The positive and negative IOC correspond to right- and left-handness of the field, respectively. 
The chirality of the system $\zeta\approx+1$ near $\mathbf{K}$ and $\zeta\approx-1$ near $\mathbf{K^{\prime}}$. 
The opposite signs of $\zeta(\mathbf{K})$ and $\zeta(\mathbf{K^{\prime}})$  can be rigorously proven through symmetry analysis. 

Given the definitions of $c$ and $\zeta$, the third term of Eq.~(\ref{phase}) is recast in $c(t')\cdot \zeta\left[\mathbf{k}(t')\right]$, explicitly encoding the instantaneous field-system chirality coupling. Then, derived from an ionization wavelet excited at an instant $t_i$ \cite{liHuygensFresnelPictureHigh2021,liHighHarmonicGeneration2023b}, the instantaneous OVSR can be assessed by the imbalance of population between $\rm K$ and $\rm K^{\prime}$ valleys (See Sec.~$\rm{I}(C)$ in SM for details \cite{hwzSM}),
\begin{equation}\label{asymmetry}
\eta_{\rm vp}=\frac{N_{c}\left(\mathbf{K}\right)-N_{c}\left(\mathbf{K}^{\prime}\right)}{N_{c}\left(\mathbf{K}\right)+N_{c}\left(\mathbf{K}^{\prime}\right)}=-\tanh{\frac{2\left[c(t_i)  \zeta_{g}\right]\bar{E}_g}{g_0}},
\end{equation}
where $N_c\left(\mathbf{K}\right)$ and $N_c\left(\mathbf{K}^{\prime}\right)$ denote the population at respective valleys and $g_0$ is a constant that describes the width of the ionization wavelet. In the above equation, $\bar{E}_g=E_{cv}[{\mathbf{K}}+\mathbf{A}(t_i)]\approx E_{cv}[\mathbf{K}^{\prime}+\mathbf{A}(t_i)]$ and $\zeta_{g}=\zeta[\mathbf{K}+\mathbf{A}(t_i)]\approx -\zeta[\mathbf{K}^{\prime}+\mathbf{A}(t_i)]$ are applied. Within this form, the sign of $c(t_i)\zeta_g$ ($+$ or $-$) gives the preference for a valley ($\rm K^{\prime}$ or  $\rm K$) contributed by the wavelet ionized at instant $t_{i}$. 

\begin{figure}[t]
\includegraphics[width=1\columnwidth]{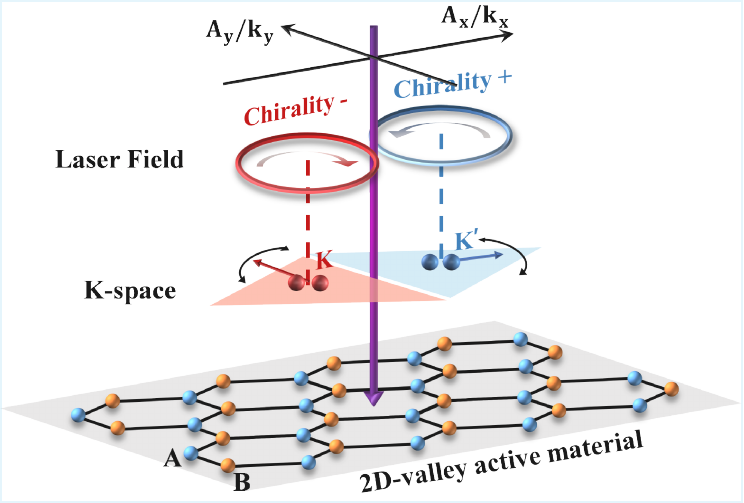}
\caption{\label{fig1}
Schematic illustration for the independent control of valley currents with chirality-separated laser fields. The closed loops represent the sub-cycle lobes of the vector potential of the driving laser. The colors indicate opposite signs of the corresponding IOC, and the offset of the loop centers shows that the lobes are biased in different directions. The opposite IOC preferentially couples to respective valleys, due to the instantaneous OVSR, while the differently biased vector potential leads to currents from each valley in different emission directions.
} 
\end{figure}

\textit{Independent control of valley currents via chirality-separated optical fields} --- The above result suggests that a single optical field with opposite IOC separated within the optical cycle, namely separated in the Lissajous figure, can be used to modulate the currents from each valley independently. As intuitively illustrated in Fig.~\ref{fig1}, the field with opposite IOC will selectively excite the $\rm K$ or $\rm K^{\prime}$ valley and then drive the excited electrons toward the respective directions of the corresponding vector potential.

To construct the required chirality-separated field, a natural starting point is to combine two circularly polarized fields with opposite helicities. However, synthesizing monochromatic circularly polarized light cannot meet the requirement due to the high symmetry of its vector potential. It is necessary to introduce a proper bias that breaks the symmetry while preserving chirality. An elementary realization of such a biased circularly polarized field is the co-rotating bicircular field (CRBF), which is composed of a circularly polarized fundamental field and its co-rotating second harmonic with the same chirality. 

\begin{figure}[t]
\includegraphics[width=1\columnwidth]{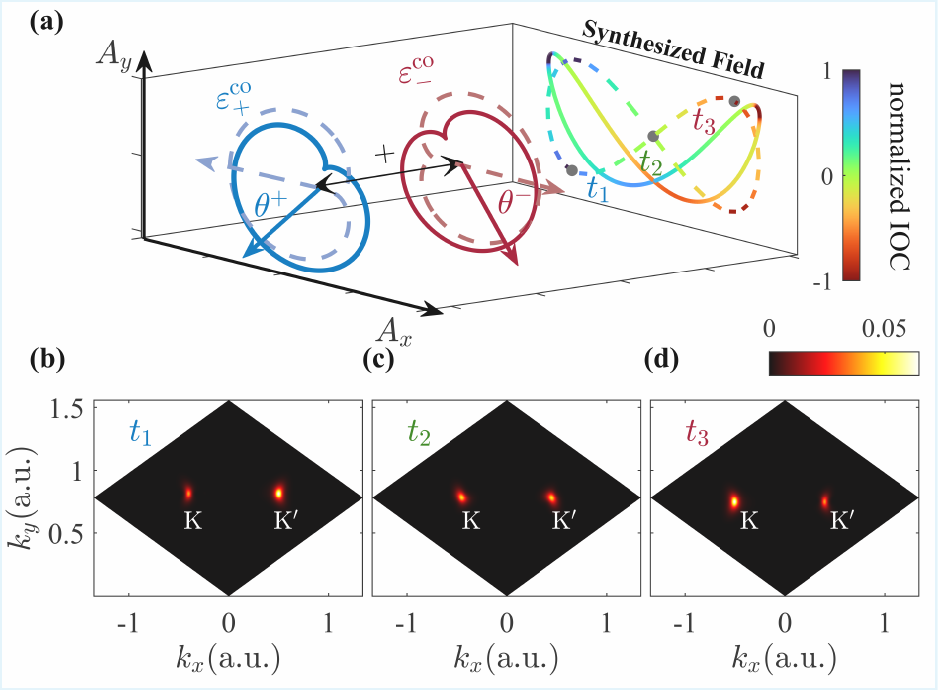}
\caption{\label{fig2}
(a) Illustration of the combination of $\boldsymbol{\varepsilon}_{+}^{\rm co}$ and $\boldsymbol{\varepsilon}_{-}^{\rm co}$, yielding chirality-separated synthesized fields. The color in the Lissajous figures indicates the normalized IOC. (b-d) The electron population contributed by the wavelets ionized at different instants marked as $t_1$, $t_2$, and $t_3$ in panel (a), respectively. 
}
\end{figure}

Analogous to the right and left circularly polarized fields, we construct two CRBFs with opposite helicities $\boldsymbol{\varepsilon}_{\pm}^{\rm co}$ as bases \cite{liHighpurityValleypolarizedCurrents2025,zhuControlGeometricPhase2022,ranaOpticalControlUltrafast2024}
\begin{equation}
    \begin{aligned}
  \boldsymbol{\varepsilon}_{\pm}^{\rm co}=\mp [ e^ {i\left(\omega t\mp\varphi_{\rm co}/2\right)} +\frac{1}{2} e^ {i\left(2\omega t\mp\varphi_{\rm co}+\phi^{\pm}\right)} ] \hat{\mathbf{e}}_{\pm},
    \end{aligned}
\end{equation}
and then the vector potential of the synthesized field is given by
\begin{equation}
   \mathbf{A}_{\rm cs}=f(t) \text{Re}\left( \sqrt{I_{\rm co}^+}\boldsymbol{\varepsilon}_{+}^{\rm co} + \sqrt{I_{\rm co}^-}\boldsymbol{\varepsilon}_{-}^{\rm co}\right).
\end{equation}
Here,  $\hat{\mathbf{e}}_\pm=\frac{1}{\sqrt2}(\hat{\mathbf{e}}_x\pm i\hat{\mathbf{e}}_y)$. $\phi^{\pm}$ indicates the phase difference between the $\omega$- and $2\omega$-components within $ \boldsymbol{\varepsilon}_{\pm}^{\rm co}$, and they determine the asymmetry direction of  $\boldsymbol{\varepsilon}_{\pm}^{\rm co}$ as depicted in Fig.~\ref{fig2}(a), respectively \cite{liHighpurityValleypolarizedCurrents2025}. $f(t)$ is the envelope of the pulse. In the case of  $\varphi_{\rm co}=\varphi-\phi^{-}+\phi^{+}$ with $\varphi$ a constant,  the pair of $ \boldsymbol{\varepsilon}_{\pm}^{\rm co}$ shares the same waveform. Consequently, the synthesized field $\mathbf{A}_{\rm cs}$ with  $I_{\rm co}^+=I_{\rm co}^-$  is non-chiral with zero net chirality averaged in a cycle, while possessing nonzero IOC with opposite signs in different sub-cycle ranges. We will focus on this kind of chirality-separated non-chiral fields in what follows. The results of varying the parameters $\varphi$ and $\phi^{\pm}$ are presented in Sec.~$\rm{II}$ in the SM \cite{hwzSM}.

Taking one of the fields as an example ($\varphi=\pi$ and $\phi^+=\phi^-=0$), we numerically demonstrate the instantaneous OVSR given by Eq.~(\ref{asymmetry}). The wavelengths of the fundamental and second-harmonic components of the field are $4000~\rm nm$ and $2000~\rm nm$, respectively. The total intensity of the synthesized laser is $5 \times 10^{10}~\text{W/cm}^2$. The laser pulse has a full duration of $12T_0$ with a trapezoidal envelope ($1T_0$  rising and falling edges and  $10T_0$ plateau), in which  $T_0=2\pi/\omega$ is the optical cycle of the fundamental field. The calculation methods and parameters are delegated in Sec.~$\rm{I(C)}$ and Sec.~$\rm{III}$ in the SM \cite{hwzSM}. We consider wavelets ionized at three representative times $t_1$, $t_2$, and $t_3$ as marked in the Lissajous figure (dashed line) in Fig.~\ref{fig2}(a). The corresponding CB population distributions are presented in Figs.~\ref{fig2}(b)-(d), respectively.  At $t_1$, the IOC is positive, i.e., $c(t_1) > 0$. The ionized electrons are primarily concentrated in the $\rm K^{\prime}$ valley, namely $N_c(\mathbf{K}^\prime)>N_c(\mathbf{K})$, as predicted by Eq.~(\ref{asymmetry}). By contrast, for $t_3$ with $c(t_3)<0$, $N_c(\mathbf{K}^\prime)<N_c(\mathbf{K})$, leading to opposite valley polarization. For $c(t_2) = 0$, identical populations are obtained at $\rm K$ and $\rm K^{\prime}$ valleys. 

\begin{figure}
\includegraphics[width=1\columnwidth]{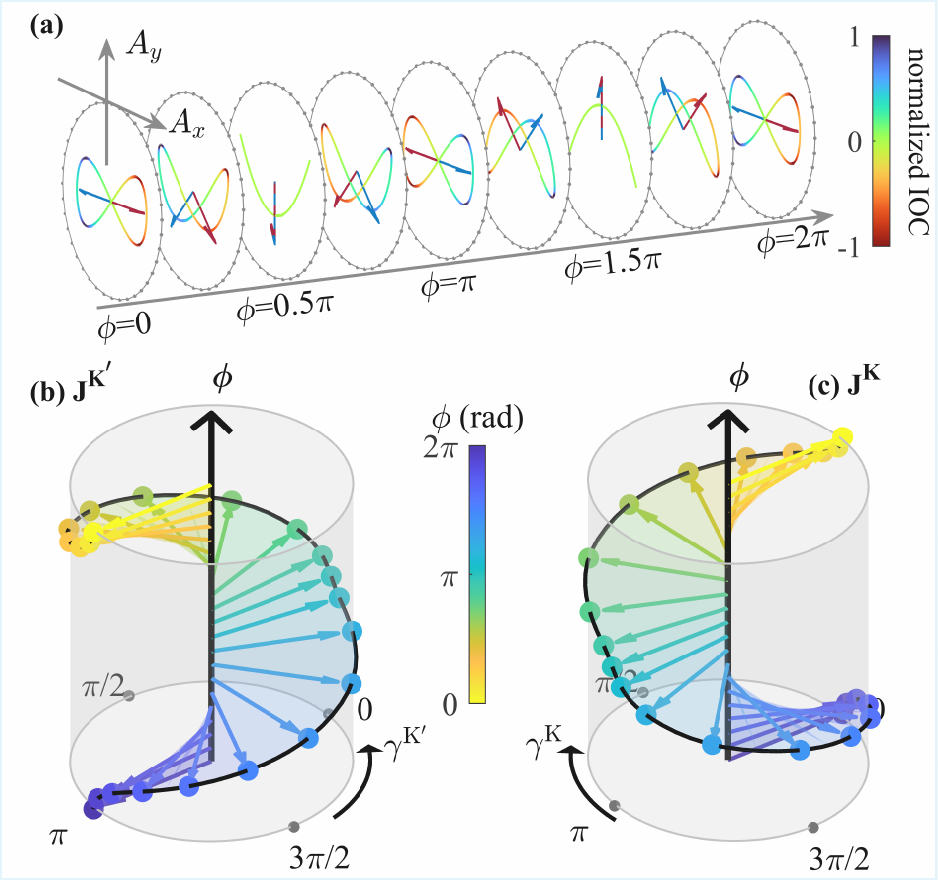}
\caption{\label{fig3}
Demonstration of the independent control over valley currents by fixing $\varphi=\pi$ and varying $\phi^+=\phi^-=\phi$. (a) The Lissajous figures of the chirality-separated fields for various $\phi$. The color reflects the normalized IOC. For $\phi=0.5\pi$ and $1.5\pi$, the laser field traces exactly the same path in both halves of the cycle. Consequently, the asymmetric directions of the vector potentials for opposite IOC align. 
(c-d) Regulation of the valley current directions by the chirality-separated fields. The arrows and dots on the cylindrical surface represent the directions of the valley currents. The black lines indicate the fitted curves of the dots, while the gradient color represents the value of $\phi$ for the optical field. 
}
\end{figure}

Then, we show the independent control of the valley currents using such chirality-separated fields. By fixing $\varphi=\pi$ and varying $\phi^+=\phi^-=\phi$, the variations of the CRBF bases are illustrated in Fig.~\ref{fig2}(a). The red and blue closed loops represent $\boldsymbol{\varepsilon}_{-}^{\rm co}$ and $\boldsymbol{\varepsilon}_{+}^{\rm co}$, respectively, with the associated arrows indicating the direction of their vector potential asymmetry $\theta^+=\pi+\phi$ and $\theta^-=-\phi$. Accordingly, the synthesized fields are shown in Fig.~\ref{fig3}(a), which exhibit a double-lobed structure. The two lobes are dominated by opposite IOC, which are encoded in red and blue, respectively.  Meanwhile, the lobes are biased to different directions, and their directions vary as $\phi$ changes. The valley currents $\mathbf{J}^{\rm K}$ and $\mathbf{J}^{\rm K^{\prime}}$ are independently rotated following the bias directions of the two lobes, as shown in Figs.~\ref{fig3}(b) and (c). Specifically, the direction of  $\mathbf{J}^{\rm K^{\prime}}$ (indicated by $\gamma^{\rm K^{\prime}}$) rotates counterclockwise with the lobe of positive IOC, while that of $\mathbf{J}^{\rm K}$ (indicated by $\gamma^{\rm K}$) follows the clockwise rotation of the lobe with negative IOC. We also demonstrate another scenario, where the valley current direction at  $\rm K$ remains almost unchanged while the current from $\rm K^{\prime}$ rotates independently to arbitrary directions (See Sec.~$\rm{IV}$ in the SM \cite{hwzSM}). 

Thus far, independent control over the valley-current direction has been established based on the above analysis. Subsequently, we showcase two key example applications, corresponding to  $\Delta\gamma=\gamma^{\rm K^{\prime}}- \gamma^{\rm K}=(n+1/2)\pi$ and $\Delta\gamma=(2n+1)\pi$ ($n\in\mathbb{Z}$), respectively.

\begin{figure}[t]
\includegraphics[width=1\columnwidth]{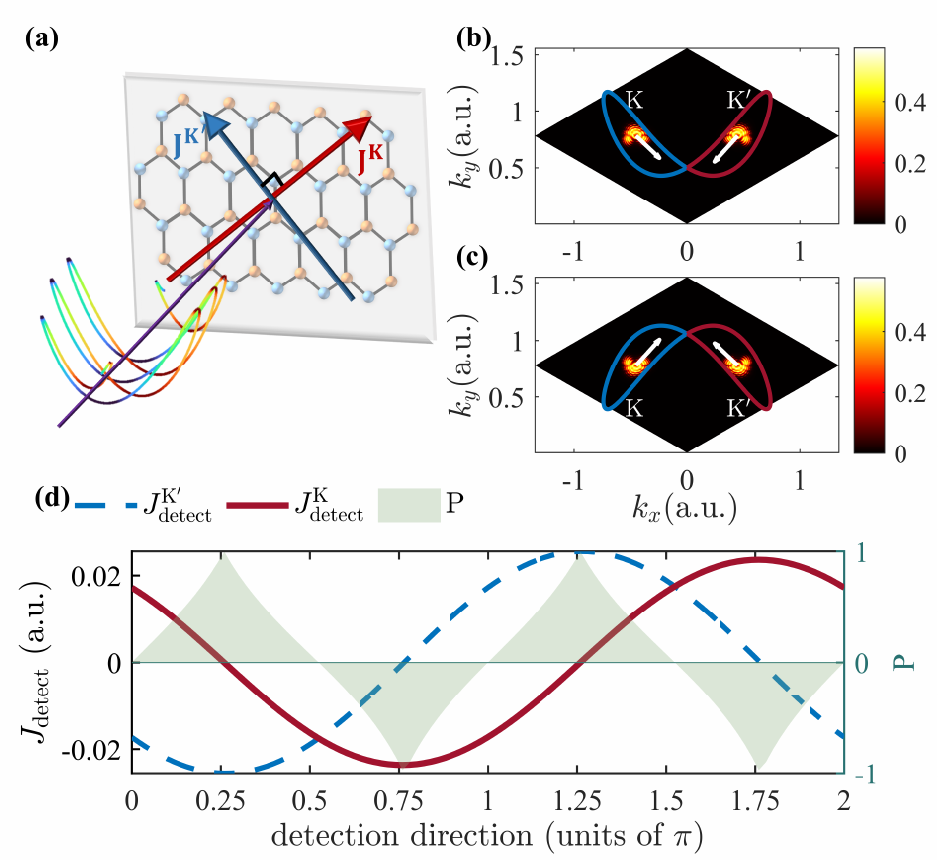}
\caption{\label{fig4}
(a) Schematic illustration of valley current separation. (b) The resultant momentum-resolved electron population and the corresponding valley current directions for the driving field of $\phi=0.4\pi$ in Fig.~\ref{fig3}. The red and blue curves are schematic diagrams for the vector potential and the sign of the IOC. (c) Same as (b), but for $\phi=1.6\pi$. (d) Projections of the currents from each valley on the detection direction and the corresponding valley purity $P$.
}
\end{figure}

\textit{Application 1: Complete valley current separation} --- As illustrated in Fig.~\ref{fig4}(a), the currents from the $\rm K$ and $\rm K^{\prime}$ valleys are emitted along mutually orthogonal directions when $\phi=0.4\pi$ or $1.6\pi$. The corresponding residual population and valley current directions are shown in Figs.~\ref{fig4}(b) and (c). This indicates that the currents from different valleys, as well as the information encoded in different valleys, can be completely separated and transferred to different detectors simultaneously.

To visualize the spatial separation of valley information, we simulate the valley currents ``detected'' in different directions by projecting the respective valley currents $\mathbf{J}^{\rm K}$ and $\mathbf{J}^{\rm K^{\prime}}$ onto the given detection direction, yielding $J^{\rm K}_{\rm{detect}}$ and $J^{\rm K^\prime}_{\rm{detect}}$ as shown in Fig.~\ref{fig4}(d). Both $J^{\rm K}_{\rm{detect}}$ and $J^{\rm K^\prime}_{\rm{detect}}$ exhibit sinusoidal variations as a function of the detection direction. Notably, when the projection of one valley current reaches its maximum, the other vanishes, demonstrating complete valley current separation. 

The complete separation of valley-specific carriers offers a highly tunable route for valley-state encoding and manipulation. In this case, high-purity valley-polarized currents will be obtained. To quantify this, we calculate the valley contribution asymmetry in the detected current as $P=\left(|J_{\rm{detect}}^{\rm K^{\prime}}|-|J_{\rm{detect}}^{\rm K}|\right)/\left(|J_{\rm{detect}}^{\rm K^{\prime}}|+|J_{\rm{detect}}^{\rm K}|\right)$, which characterizes the purity of valley-polarized current. As shown by the green area in  Fig.~\ref{fig4}(d), the absolute value $|P|$ reaches its maximum of 1 at angles  $(0.25+n/2)\pi$ ($n\in\mathbb{Z}$),  indicating the generation of valley-polarized current with 100\% purity. Moreover, the $\pi/2$ spacing between $P = \pm1$ points confirms the orthogonality of the two valley currents. Compared with existing approaches that rely on unbalanced excitation in different valleys,  where the excitation asymmetry often falls short of realizing 100\% selectivity \cite{oliaeimotlaghFemtosecondValleyPolarization2018,neufeldLightDrivenExtremelyNonlinear2021,motlaghAnomalousUltrafastAlloptical2021,ranaOpticalControlUltrafast2024,sharmaDirectCouplingLight2024a,liHighpurityValleypolarizedCurrents2025,leskoOpticalControlElectrons2025a,gillCreationControlValley2025}, our scheme provides a new way to obtain a high-purity valley-polarized current.

\begin{figure}
\includegraphics[width=1\columnwidth]{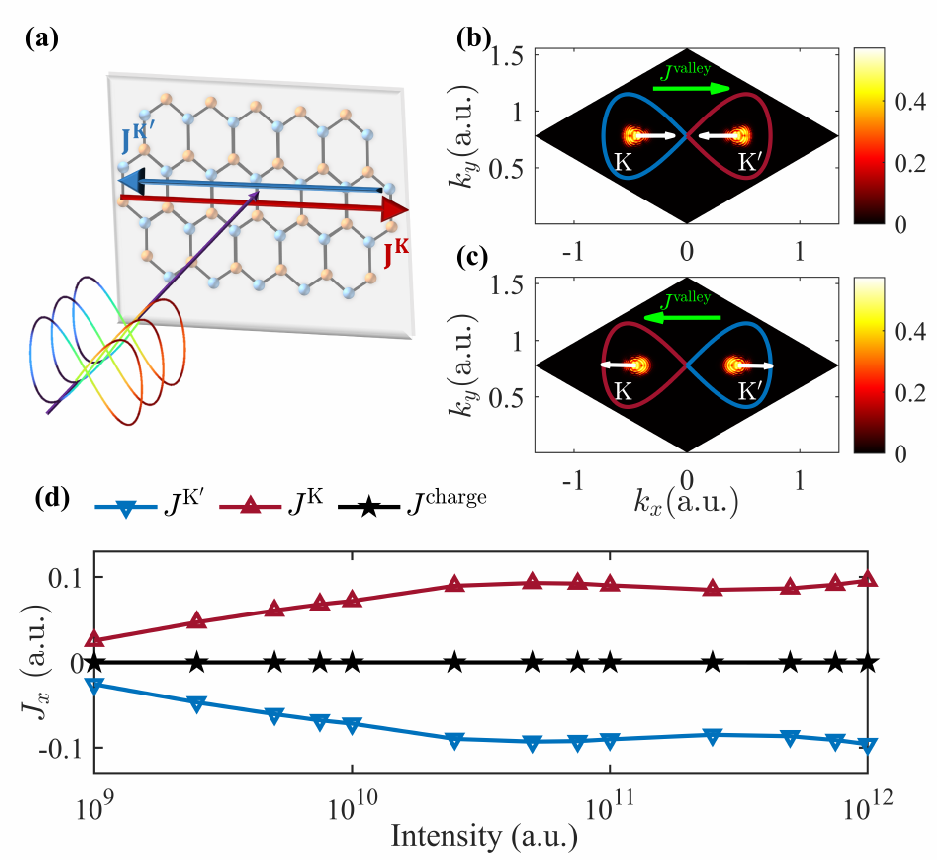}
\caption{\label{fig5}
(a) Schematic illustration of pure valley current generation. (b) The resultant momentum-resolved electron population and the corresponding valley current directions for the driving field of $\phi=0$ in Fig.~\ref{fig3}. The green arrow indicates the direction of the pure valley current (transport of pseudo-spin). (c) Same as (b), but for $\phi=\pi$. (d) Magnitudes of the currents from each valley and the total charge current as a function of total laser intensity.
}
\end{figure}

\textit{Application 2: Generation of pure valley current} --- As illustrated in Fig.~\ref{fig5}(a), equal currents from the $\rm K$ and $\rm K^{\prime}$ valleys are emitted along opposite directions when $\phi=0$ or $\pi$, yielding a pure valley current with zero net charge flow. The corresponding residual populations for these driving fields are shown in Figs.~\ref{fig5}(b) and (c). Around  $\rm K$ and $\rm K^{\prime}$, the population exhibits a pronounced directional asymmetry, which gives rise to a finite current in each valley. Such asymmetry originates from the asymmetric sub-cycle waveform of the vector potential for opposite IOC. Meanwhile, the population remains symmetric across the BZ because the whole laser field possesses mirror symmetry, constraining the charge current to zero. It is worth noting that the Lissajous figures of the vector potentials are exactly the same in Figs.~\ref{fig5}(b) and (c), but the results are essentially different. The directions of the valley currents  $\mathbf{J}^{\rm K}$ and $\mathbf{J}^{\rm K^\prime}$ are reversed, leading to opposite transport directions of the valley information. This further highlights the determinant role of the IOC in our scheme.

As zero charge current is guaranteed by the symmetry, the present approach to producing pure valley current is robust against fluctuations of laser parameters. Figure~\ref{fig5}(d) presents the $x$ components of $\mathbf{J}^{\rm K}$, $\mathbf{J}^{\rm K^\prime}$, and charge current $\mathbf{J}^{\rm charge}=\mathbf{J}^{\rm K}+\mathbf{J}^{\rm K^{\prime}}$ as the total intensity of the driving laser increases from $1\times 10^9$ to $1\times 10^{12}~\rm W/cm^2$ for $\phi=0$. The $y$ components of the currents are zero due to the mirror symmetry of the field about the $x$-axis.
One can see that the magnitudes of $J_x^{\rm K}$ and $J_x^{\rm K^{\prime}}$ increase with the increase of the laser intensity. In contrast, the corresponding charge current remains at a negligible level, indicating the absence of net charge transport. Furthermore, we calculate $Q=\left(|\mathbf{J}^{\text{valley}}|-|\mathbf{J}^{\rm charge}|\right)/\left(|\mathbf{J}^{\text{valley}}|+|\mathbf{J}^{\rm charge}|\right)$, where  $\mathbf{J}^{\text{valley}}=\mathbf{J}^{\rm K}-\mathbf{J}^{\rm K^{\prime}}$ represents the current component associated with the VDOF. When $Q=1$, the total current is a pure valley current where no net charge is transported. In contrast, when $Q=-1$, the total current is a complete charge current without transport of valley information. 
Our calculations find that under the laser intensities used in Fig.~\ref{fig5}(d), $Q$ consistently equals 1 up to the fourth decimal place.

Pure valley current, closely resembling pure spin current, is of significant interest for low-dissipation information processing and valleytronics-based logic devices \cite{makValleyHallEffect2014a,gorbachevDetectingTopologicalCurrents2014a,shanOpticalGenerationDetection2015a,suiGatetunableTopologicalValley2015a,shimazakiGenerationDetectionPure2015a,gillUltrafastAllopticalGeneration2025a}. Our scheme achieves robust and efficient generation of pure valley current within a single optical field, enabling ultrafast control of the charge-neutral information transport on the optical cycle timescale. 

\textit{Conclusion}---In summary, we reveal the instantaneous OVSR that clarifies the role of IOC in valley-resolved excitation. Building on these insights, we propose and demonstrate a scheme with chirality-separated optical fields to achieve independent control over valley currents. Utilizing this scheme, one can fully utilize the two VDOFs, realizing the on-demand design of the valley currents, thereby fulfilling the multifarious demands inherent to the realm of valleytronics. Our work presents a general and robust all-optical method for manipulating valley currents, offering new possibilities for ultrafast valleytronic and quantum information processing. 


\textit{Acknowledgement}---This work was supported by National Key Research and Development Program under grant No. 2023YFA1406800, the National Natural Science Foundation of China under grants Nos. 12174134, 12450406, 12225406, and 12021004. The computation is completed in the HPC Platform of Huazhong University of Science and Technology.

\bibliography{background}

\begin{thebibliography}{54}
\bibitem [{not()}]{note_c}%
  \BibitemOpen
  \href@noop {} {\bibinfo {title} {The definition in {{Eq}}.~(2) differs from that given in the references by a factor of the electric field amplitude.}}\BibitemShut {Stop}%
\end{thebibliography}
\end{document}